


\def\sqr#1#2{{\vcenter{\hrule height.#2pt
             \hbox{\vrule width.#2pt height#1pt \kern#1pt
             \vrule width.#2pt}
             \hrule height.#2pt}}}


\def\pmb#1{\setbox0=\hbox{#1}\kern-.025em
    \copy0\kern-\wd0\kern.05em\kern-.025em\raise.029em\box0}

\def\a{\alpha}         
\def\b{\beta}       \def\m{\mu}
\def\g{\gamma}      \def\G{\Gamma}    \def\n{\nu}
\def\d{\delta}      
                      
\def\ve{\varepsilon}                  \def\s{\sigma}  \def\S{\Sigma}
                         \def\t{\tau}
        \def\f{\varphi}
                   \def\o{\omega}  \def\O{\Omega}
\def\k{\kappa}

\def\p{\partial}
\def\du{{}^{\ast}\! }

\def\cH{{\cal H}}


\def\fr#1 #2{\hbox{${#1\over #2}$}}        
\def\leaderfill{\leaders\hbox to 1em{\hss.\hss}\hfill}

\def\section#1{
  \vskip.7cm\goodbreak
  \noindent{\bf #1}
  \nobreak\vskip.4cm\nobreak  }

\def\subsection#1{
  \vskip.3cm\goodbreak
  \noindent{\it #1}
  \vskip.3cm\nobreak}

\def\ver#1{\left\vert\vbox to #1mm{}\right.}



\font\eightrm=cmr8                                
\font\twelvebf=cmbx12                             

\def\preprint#1{\hskip10cm #1 \par}
\def\date#1{\hskip10cm #1 \vskip2cm}
\def\title#1{ \centerline{\twelvebf\uppercase{#1}} }
\def\titlef#1{ \vskip.2cm                          
      \centerline{ \tenbf\uppercase{#1}
                   \hskip-5pt ${\phantom{\ver{3}}}^\star$  }
      \vfootnote{$^\star$}{\eightrm Work supported in part
      by the Serbian Reserch Foundation, Yugoslavia.} \vskip.5cm }
\def\author#1{ \vskip.5cm \centerline{#1} }
\def\institution#1{ \centerline{\it #1} }
\def\abstract#1{ \vskip3cm \centerline{\bf Abstract}
                 \vskip.3cm {#1} \vfill\eject }


\magnification=\magstep1
\null

\preprint{IF- 3/95}
\date{March  1995}

\title{CHIRAL SYMMETRIES OF THE WZNW MODEL}
\titlef{BY HAMILTONIAN METHODS}
\vskip1cm
\author{B. Sazdovi\'c}
\institution{Institute of Physics, P.O.Box 57,11001 Beograd, Yugoslavia}
\abstract{
The connection between the Kac-Moody algebras of currents and the
chiral symmetries of the two dimensional WZNW model is clarified.
It is shown that only the zero modes of the
Kac-Moody currents are the first class constraints, and that, consequently,
the corresponding gauge symmetries are chiral. }

\section{1. Introduction}
It is well known that the two dimensional Wess-Zumino-Novikov-Witten
(WZNW) model [1-3]  in addition to the  conformal
invariance possesses an invariance under the {\it chiral} gauge
transformations
$$
g^{\O_+ \O_-}(x^+,x^-)= \O_+(x^+)g(x^+,x^-)\O_-^{-1}(x^-)     \eqno(1)
$$
when $\O_+$ and $\O_-$ are arbitrary group valued functions of the light cone
coordinates
$$
x^{\pm}={1\over \sqrt{2}}(x^0 \pm x^1).      \eqno(2)
$$

It is common in the literature [3-5] to connect the
Kac-Moody (KM) currents with the symmetry (1). The goal of this paper
is to clarify this connection from the point of view of Dirac's
method of constraints [6]. The generators of the symmetry must
be the first-class (FC) constraints while "almost all" KM currents
are the second-class (SC) constraints. We are going to
solve this puzzle by showing that only one Fourier mode of the
KM currents is the FC constraint,
and this mode is the generator of the chiral symmetry (1).
This explains why the symmetry is chiral.

There is another nonstandard application of this canonical analysis.
Usualy we use this approach to obtain {\it local} gauge symmetries
(the parameters of the transformations depend on all space-time
coordinates). In this example we are able to find {\it chiral}
gauge symmetries (the parameters of the transformations depend only
on some particular combinations of the coordinates, here $x^+$ and
$x^-$). The FC constraints as the generators of these transformations
are the Fourier components of the complete constraints.

\section{2. The two dimensional WZNW model}

We will consider the WZNW model described by the action [3]
$$
S(g)=S_0(g)+k\G(g).   \eqno (3a)
$$
Here
$$
S_0(g)={k\over 16\pi} \int\limits_{\Sigma} (\du v_-,v_-) \eqno(4a)
$$
is the action of the non linear $\s$-model on the field $g: \S \to G$
where $\S$ is two dimensional space-time and $G$ is a semisimple Lie group.
The expression $(X,Y)$ is the Cartan-Killing form, invariant under the adjoint
action of the group element $\O \in G$
$$
(\O X \O^{-1},\O Y \O^{-1})=(X,Y),    \eqno(5)
$$

$$
v_-=g^{-1}dg \qquad v_+=gdg^{-1}=-dgg^{-1}   \eqno(6)
$$
are the left (right) invariant Maurer-Cartan forms and
$\du v=\du d x^\mu v_\mu =\ve^\mu {}_\nu d x^\nu v_\mu$ is the dual of $v$.
The second term
$$
\Gamma(g)={1\over 24\pi}\int\limits_B {1\over2} (v_-,[v_-,v_-])  \eqno(7a)
$$
is the Wess-Zumino topological term where the integration is carried over a
three dimensional manifold $B$, whose boundary is $\S$  $(\S=\p B)$.

Instead of working with the field $g(x)$ we will benefit from taking
local coordinates on the group manifold $\f^\a (x)$ as  canonical
coordinates.
Let $t_a$ be the generators of the Lie algebra of $G$ satisfying
 $[t_a,t_b]=f_{ab}{}^c t_c$, and let $\g_{ab}=(t_a,t_b)=f_{ac}{}^df_{bd}{}^c$
be the group metric. The coordinate indices $\a,\b,\g,...$ and the
Lie algebra indices $a,b,c,...$ run over the same range.

We define the veilbeins on $G$, $E_\a{}^a(\f)$, as the coeficients of the
expansion of $v$ in $d\f^\a$ and $t_a$,
$$
v_A=d\f^\a E_{A\a}{}^a(\f)t_a=d\f^\a E_{A\a}    \eqno(8)
$$
where $A={+,-}$ are light cone indices. There are two
veilbeins, $E_{+\a}{}^a$ and $E_{-\a}{}^a$,  as well as
two Maurer-Cartan forms, $v_+$ and $v_-$.
The metric on the group manifold
$$
H_{\a\b}(\f)=(E_{A\a},E_{A\b})=\g_{ab}E_{A\a}{}^aE_{A\b}{}^b   \eqno(9)
$$
(no sumation over A) does not depend on $A$ ($H_+=H_-$) as a consequence
 of the relation $v_-=-g^{-1}v_+ g$ and the property (5) of the
Cartan-Killing form.

The action (4a) can be now rewritten in the form
$$
S_0(\f)={k \over 8\pi} \int\limits_{\S} d^2x ({\fr -1 2})
\eta^{\m\n}H_{\a\b}(\f)
\p_{\m}\f^\a \p_\n\f^\b.    \eqno(4b)
$$

Since the three form $v^3$ is a closed form on the three dimensional manifold
$B$,
we have locally
$$
{\fr 1 2}(v_A,[v_A,v_A])=-3!d\t_A(\f)   \eqno(10)
$$
where $\t_A$ is a 2-form
$$
\t_A(\f)={\fr 1 2}\t_{A\a\b}d\f^\a d\f^\b.  \eqno(11)
$$
Note that $\t_-=-\t_+$.Using the Stoke's theorem the Wess-Zumino term
$\G$ can be locally converted to an integral over the space time $\S$
$$
\G(\f)=-{1\over 4\pi}\int\limits_{\S=\p B}\t_- =
{1\over 8\pi}\int\limits_\S d^2x \ve^{\mu\nu} \t_{-\a\b}(\f)
\p_{\mu} \f^\a \p_{\nu} \f^\b .   \eqno(7b)
$$

The local expression for the action
$$
S(\f)=S_0(\f)+k\G(\f)=
\k \int\limits_\S d^2x [-{\fr 1 2} \eta^{\mu\nu} H_{\a\b}(\f) +
 \ve^{\mu\nu} \t_{-\a\b}(\f)]\p_{\mu}\f^\a \p_{\nu}\f^\b     \eqno(3b)
$$
writen in the light cone coordinates takes the form
$$
S(\f)=-\k \int\limits_\S d^2x [H_{\a\b}(\f)+2\t_{-\a\b}(\f)] \p_- \f^\a
\p_+\f^\b          \eqno(3c)
$$
where $\k\equiv {k\over 8\pi}$,$\quad \p_{\pm}={1\over
\sqrt{2}}(\p_0\pm \p_1) $ and $\ve^{01}=-1$. The action in the form (3c)
is very convenient for the canonical analisys. For the final
conclusion we will need boundary conditions, as will be discussed later.

\section{3. Constraints}

We are going to use Dirac's canonical approach to find the gauge
symmetries. The hamiltonian equations of motion describe
dinamical evolution of a system by a trajectory in a phase space.
The presence of an arbitrary multiplier in the hamiltonian
means that for given initial conditions we will not obtain
the unique solution, one trajectory, but a whole set of
trajectories, describing the same physical state. The FC constraints
are generators of the mapping between these trajectories that do not
change the phisical state. These unphysical transformations are
the gauge symmetries of the hamiltonian equations of motion.
The method essentialy
depends on the choice of time variable. If we take $x^0$ as a time
coordinate and use the bilinear expression in the time derivative
(3b), as a action, there are no
constraints and consequently we can not find the symmetries this way.

The expression (3c) for the action is more promising because it is linear
in the $\p_- \f$ and $\p_+ \f$. Taking first $x^-$ and then $x^+$ as
time coordinates we will find some primary constraints which give us a
hope to find some symmetries of the system. To treat both cases
together we will use the light cone indices $A=\{+,-\}$ with
$x^A=\{x^+,x^-\}$ and $ \p_A=\{\p_+,\p_-\}$. These two cases are
independent and there are no summations over repeated indices $A$.

In the first case  $(A=-)$ we will take $x^-=\t$ and
$x^+=\s$ and in the second one $(A=+)$  we will take
$x^+=\t$ and $x^-=-\s$. (The minus sign we adopt to preserve the
orientation between coordinate axis.)

The action corresponding to the choice $x^A=\t$ has the form
$$
S_A(\f)=\k \int d^2x[(-1)^A H_{\a\b}-2\t_{-\a\b}]{\dot \f}^\a
\f^{\prime \b}.     \eqno(12)
$$
The canonical momentum conjugate to the variable $\f^\a$ is
$$
\pi_\a=\k [(-1)^A H_{\a\b} -2\t_{-\a\b}]\f^{\prime \b}
$$
and therefore, there are primary constraints:
$$
J_{A\a} =\pi_\a+2\k \t_{-\a\b}\f^{\prime \b}-(-1)^A\k
H_{\a\b}\f^{\prime \b}.   \eqno(13)
$$
We prefer to rewrite these constraints with the Lie algebra indices as
$$
J_{Aa}=-E_{Aa}{}^\a J_{A\a}     \eqno(14)
$$
where $E_{Aa}{}^\a$ is the inverse of $E_{A\a}{}^a$.
The Poisson brackets (PB) between these constraints define two
independent KM algebras ($A=+ \quad or \quad -$):
$$
\{J_{Aa}(x),J_{Ab}(y)\}=f_{ab}{}^c J_{Ac}(x)\d(\s_x-\s_y)
-(-1)^A 2\k \g_{ab}\d^{\prime}(\s_x-\s_y).  \eqno(15a)
$$
The expression (14) is a convenient
linear combination of the previous constraints. We adapt a
correspondence between $J_\pm$ and $E_\pm$  in
order to obtain the algebra (15a).

The last term in the KM algebra contains the derivative of the
$\d$-function and is known as a Schwinger term. In the two dimensional
theory with free massless fermions this term originates from the
anomalous commutator algebra [3] and has a quantum character.
It drastically changes the nature of the constraints [7] from the
FC constraints to the SC ones. In the WZNW model the Schwinger term arose
classically in the PB algebra, and it has important implications
for the nature of the classical symmetries of the theory,

The action is linear in the time derivative and consequently the
canonical hamiltonian density is zero, $\cH_c =0$, while the total
hamiltonian takes the form
$$
H_{AT}=\int d\s \cH_{AT}=\int d\s u_A{}^a J_{Aa}.      \eqno(16a)
$$

The consistency condition for $J_{Aa}$ leads to
$$
{\dot J}_{Aa}(x)=\{J_{Aa}(x),H_{AT}\} \approx -(-1)^A 2\k
u^{\prime}{}_{Aa} (x) \approx 0.    \eqno(17a)
$$

The multipliers in (17a) looks like
determined $u^{\prime}{}_{Aa}=0$ which means that $J_{Aa}$ should
be SC constraints. More precisely, only the first derivative of the
multipliers is determined and the corresponding constraints are the SC.
The $\s$-independent part is arbitrary function of $\t$ and corresponding
constraints are the FC. The same conclusion we can obtain from eq.(15a).
The Schwinger term contains the derivative of the $\d$-function
and gives zero on the $\s$-independent functions. It means that
the Schwinger term does not give a contributions for some modes of the
currents and these modes are the FC constraints.

Let us make a more precise investigation.
We assume that space is compact which means that for all variables
$X=\{J_{Aa}, u_A{}^a,...\}$ we have  $X(\t,\s+L)=X(\t,\s)$. We will use
the Fourier expansion
$$
X(x)={1\over L}\sum_{n \in Z} X_n(\t_x)e^{-in\o \s_x}, \qquad
X_n(\t_x)=\int\limits_0^L d\s_x X(x)e^{in\o \s_x}  \eqno(18)
$$
where $\o={2\pi \over L}$, and the $\d$-function representation
$$
\d(\s)={1\over L} \sum_{n\in Z} e^{-in \o \s}.   \eqno(19)
$$
The PB between modes $J_{Aan}(\t)$ are
$$
\{J_{Aan},J_{Abm}\}=f_{ab}{}^c J_{Ac(n+m)}-(-1)^A
4\pi i\k n\g_{ab} \d_{n+m,0}    \eqno(15b)
$$
while plus and minus constraints commute.
This expression does not depend on the period $L$.
In terms of modes the total hamiltonian becomes
$$
H_{AT}(\t_x)=\int\limits_0^Ld\s_x\cH_{AT}(x)=
{1\over L}\sum_{n\in Z}u_A{}^a{}_n(\t_x) J_{Aa(-n)}(\t_x).  \eqno(16b)
$$

Now it is easy to separate the FC and the SC constraints, and
check the consistency
condition for the primary constraints $J_{Aan}(\t)$:
$$
{\dot J}_{Aan}(\t)=\{J_{Aan}(\t),H_{AT}\} \approx
-(-1)^A{4\pi i \over L}\k n u_{Aan}(\t) \approx 0 .    \eqno(17b)
$$
For $n=0$ the multipliers $u_{Aa0}(\t)$ are arbitrary functions
of time and  $J_{Aa0}(\t)$ constraints are FC.
For $n\not= 0$ we can determined the multipliers
$u_{Aan}(\t)=0$ $(n \not= 0)$  which means that $J_{Aan}$ $(n\not= 0)$
are SC constraints. The same conclusion follows from (15b) because
$$
\eqalign{&
\{J_{Aa0},J_{Abn}\}=f_{ab}{}^c J_{Acn} \approx 0,  \cr
&
\{J_{Aan},J_{Ab(-n)}\} \approx
-(-1)^A 4 \pi i \k n \g_{ab} \not= 0 \quad for \quad n \not= 0. \cr}  \eqno(20)
$$

In any case there are no secondary constraints. The total hamiltonian
is now
$$
H_{AT}={1 \over L} u_A{}^a{}_0(\t) J_{Aa0}(\t).  \eqno(16c)
$$
In sec.5 we will verify that expression (16c) gives the correct
equations of motion.

The assumption of the periodicaly boundary conditions is not necessary.
The general solution of (17a) is $u_{Aa}(\s,\t)=u_{Aa}(\t)$ and substituting
this into (16a) we obtain (16c) with $u_A{}^a(\t)$ instead  of
${\fr 1 L} u_A{}^a{}_0(\t)$
and $J_{Aa0} = \int d \s J_{Aa}(\s,\t)$. The constraints $J_{Aa0}$ are
FC because they correspond to the arbitrary multiplier $u_{Aa}(\t)$.
They also weakly vanish with all constraints, which can be easily
verified integrating eq.(15a) over $\s_x$ . For our purpose any boundary
condition with finite $J_{Aa0}$ is a good condition.

\section{4. Chiral gauge symmetries}

The presence of arbitrary multipliers  $u_A{}^a{}_0(\t)$
in the total hamiltonian means that the theory  has a gauge symmetry.
We will use Castellani's method [8]  to obtain the  gauge generators
and  the corresponding gauge transformations.The gauge parameters
$\o_A{}^a(\t)$ will depend only on the {\it time} coordinate $\t$,
as well as the FC constraints $J_{Aa0}$ and the arbitrary
multipliers $u_A{}^a{}_0$.

All the FC constraints  $J_{Aa0}$ are primary in this case, and the
expression for the gauge symmetry generator is
$$
G_A=\o_A{}^a(\t)J_{Aa0}(\t).  \qquad (A=+,-) \eqno(21)
$$
The gauge transformations produced by $G_A$ are
$$
\d_A \f^\a=\{\f^\a,G_A \}=-\o_A{}^a(\t) E_{Aa}{}^\a .  \eqno(22)
$$

We are interested in the gauge transformations of the field $g$. Using
eqs.(6) and (8) we can connect the transformations of the fields $g$
and $\f$
$$
g^{-1}\d_-g=\d_-\f^\a E_{-\a} {}^a t_a    \qquad
-\d_+gg^{-1}=\d_+ \f^\a E_{+\a} {}^a t_a.    \eqno(23)
$$
With the help of eq.(22) we get
$$
\d_-g=-g\o_-    \qquad     \d_+g=\o_+ g   \eqno(24)
$$
where $\o_A=\o_A{}^a t_a$.

Because $\o_A$ is infinitesimal we can write
$$
g^{\o_-}=g+\d_-g=g(1-\o_-) \approx ge^{-\o_-}    \qquad
g^{\o_+}=g+\d_+g=(1+\o_+)g \approx e^{\o_+}g.  \eqno(25)
$$

Recalling that $\o_-\equiv \o_-(x^-)$ and $\o_+\equiv \o_+(x^+)$
we have
$$
g^{\O_-}=g \O^{-1}_-(x^-)  \qquad    g^{\O_+}=\O_+(x^+)g  \eqno(26)
$$
where $\O_A(x^A)=e^{\o_A(x^A)}$. Equations (26) are equivalent with
the equation (1). Therefore we demonstrated that the gauge symmetries
generated by the zero modes of the KM currents are exactly the
known chiral gauge symmetries.

\section{5. Equations of motion}

For completeness we briefly derive the hamiltonian equations of motion.
For the fields $\f^\a$ we have
$$
{\dot \f}^\a=\{\f^\a,H_{AT}\}=-u_A{}^a{}_0 E_{Aa}{}^\a .   \eqno(27)
$$
The equations of motion for the field $g$ are obtained from eqs.(6)and (8)
in terms of the equations for $\f^\a$
$$
g^{-1}{\dot g}={\dot \f}^\a E_{-\a}{}^at_a    \qquad
-{\dot g}g^{-1}={\dot \f}^\a E_{+\a}{}^at_a     \eqno(28)
$$
and by using eq.(27) we find
$$
g^{-1}{\dot g}=-u_{-0}(\t)   \qquad    {\dot g}g^{-1}=u_{+0}(\t)   \eqno(29a)
$$
where $u_{A0}=u_A{}^a{}_0 t_a$.

If we now recall our definition of time $\t=x^A$, it follows that
${\dot g}=\p_-g$ for $A=-$ and  ${\dot g}=\p_+g$  for $A=+$,
so that eqs.(29a) can be rewritten as
$$
g^{-1}\p_-g=-u_{-0}(x^-)  \qquad  \p_+gg^{-1}=u_{+0}(x^+).   \eqno(29b)
$$
Therefore we obtain the well known equations of motion,
$$
\p_+(g^{-1}\p_-g)=0   \qquad     \p_-(\p_+gg^{-1})=0     \eqno(30)
$$
which are, as well as the equations (29b), evidently invariant
under the chiral gauge transformations (1).

\section{6. Conclusion}

The canonical formalism in the light cone coordinates is developed for the
two dimensional WZNW model. Using $x^-$ and $x^+$
as the evolution parameters we obtain the two commuting KM algebras
of constraints. We made a mode expansion in order to
separate the FC constraints from the SC ones. Only the zero modes
are FC constraints and they are the
generators of the gauge symmetries. The only arbitrary parameters
are zero mode gauge parameters. They depend on the $x^-$ in the
first case and on the  $x^+$ in the second one, and this explains the
chirality of the gauge symmetry.

The connection between KM algebra and chiral gauge symmetry is now
clear. The KM algebra contains infinitely many modes of the SC
constraints (it looks like the algebra of SC constraints) and
only one FC mode which is the generator of the symmerty (1).

This result is obtained by the application of the canonical approach to
a new kind of symmetry when the gauge parameter does not  depend
on all coordinates, but only on some particular
combinations of the coordinates [9].

\section{References}

\item{[1]} J. Wess and B. Zumino, Phys.Lett. {\bf 37B} (1971) 95.

\item{[2]} S. P. Novikov, Sov.Math.Doklady {\bf 24} (1981) 222;
Usp.Math.Nauk. {\bf 37} (1982) 3.

\item{[3]} E. Witten, Nucl.Phys. {\bf B223} (1983) 422;
Commun.Math,Phys. {\bf 92} (1984) 455.

\item{[4]} V. G. Knizhnik and A. B. Zamolodchikov, Nucl.Phys.
{\bf B247} (1984) 83.

\item{[5]} D. Gepner and E. Witten, Nucl.Phys. {\bf B278} (1986) 493.

\item{[6]} P. A. M. Dirac, Lectures on quantum mechanics, Belfer Graduate
School of Science (Yeshiva Univ., 1964).

\item{[7]} L. D. Faddeev, Phys.Lett. {\bf 145B} (1984) 81;

L. D. Faddeev and S. L. Shatashvili, Phys.Lett. {\bf 167B} (1986) 225.

\item{[8]} L. Castellani, Ann.Phys. (New York) {\bf 143} (1982) 357.

\item{[9]} M. Blagojevi\'c, M. Vasili\'c and T. Vuka\v sinac,
Class.Quan.Grav. {\bf 11} (1994) 2134.

\end